\documentclass[12pt,preprint]{aastex}
\usepackage{epsfig}

\def \kms{\ifmmode{~{\rm km\,s}^{-1}}\else{~km~s$^{-1}$}\fi}
\def \vhel{\ifmmode{V_{{\rm hel}}}\else{$V_{{\rm hel}}$}\fi}
\def \vsys{\ifmmode{V_{{\rm sys}}}\else{$V_{{\rm sys}}$}\fi}
\def \vrad{\ifmmode{V_{{\rm rad}}}\else{$V_{{\rm rad}}$}\fi}
\def \vexp{\ifmmode{V_{{\rm exp}}}\else{$V_{{\rm exp}}$}\fi}
\def \degree{\ifmmode{^{\circ}}\else{$^{\circ}$}\fi}
\def \lsun{\ifmmode{{\rm\ L}_\odot}\else{${\rm\ L}_\odot $}\fi}
\def \msun{\ifmmode{{\rm\ M}_\odot}\else{${\rm\ M}_\odot$}\fi}
\def \myr{\ifmmode{{\rm\ M}_\odot{\rm\ yr}^{-1}}\else{${\rm\ M}_\odot$ yr$^{-1}$
}\fi}
\def \teff{\ifmmode{{\rm{T}}_{\rm eff}}\else{${\rm{T}}_{\rm eff}$}\fi}
\def \mdot{\ifmmode{{\rm\dot{M}}}\else{${\rm\dot{M}}$}\fi}

\newcommand{\ha}{H$\alpha$}

\newcommand{\oiii}{[O\,{\sc iii}]}
\newcommand{\oiiil}{[O\,{\sc iii}]\ $\lambda$5007\,\AA}

\newcommand{\nii}{[N\,{\sc ii}]}

\newcommand{\nev}{[Ne\,{\sc v}]}
\newcommand{\nevl}{[Ne\,{\sc v}]\ $\lambda$3426\,\AA}

\def \st{\ifmmode{^{\mathrm{st}}}\else{${^{\mathrm{st}}}$}\fi}
\def \nd{\ifmmode{^{\mathrm{nd}}}\else{${^{\mathrm{nd}}}$}\fi}
\def \rd{\ifmmode{^{\mathrm{rd}}}\else{${^{\mathrm{rd}}}$}\fi}
\def \th{\ifmmode{^{\mathrm{th}}}\else{${^{\mathrm{th}}}$}\fi}

\shorttitle{HST Observations of RS Ophiuchi}
\shortauthors{Bode et al.}

\begin{document}


\title{Hubble Space Telescope Imaging of the Expanding Nebular Remnant of the Recurrent Nova RS Ophiuchi (2006)}


\author{M.F. Bode and D.J. Harman}
\affil{Astrophysics Research Institute, Liverpool John Moores University, Birkenhead, CH41 1LD, UK}
\email{mfb@astro.livjm.ac.uk, dh@astro.livjm.ac.uk}

\author{T.J. O'Brien}
\affil{Jodrell Bank Observatory, School of Physics and Astronomy, University of Manchester, Macclesfield, SK11 9DL, UK}
\email{tim.obrien@manchester.ac.uk}

\author{Howard E. Bond}
\affil{Space Telescope Science Institute, 3700 San Martin Drive, Baltimore, MD 21218, USA}
\email{bond@stsci.edu}

\author{S. Starrfield}
\affil{School of Earth and Space Exploration, Arizona State University, P.O. Box 871404,Tempe, AZ 85287-1404, USA}
\email{sumner.starrfield@asu.edu}


\author{M.J. Darnley}
\affil{Astrophysics Research Institute, Liverpool John Moores University, Birkenhead, CH41 1LD, UK}
\email{mjd@astro.livjm.ac.uk}

\author{A. Evans} 
\affil{Astronomy Group, School of Physical and Geographical Sciences, Keele University, ST5 5BG, UK}
\email{ae@astro.keele.ac.uk} 

\author{S.P.S. Eyres} 
\affil{Centre for Astrophysics, University of Central Lancashire, Preston, PR1 2HE, UK}
\email{spseyres@uclan.ac.uk} 




\begin{abstract}

We report Hubble Space Telescope imaging obtained 155 days after the 2006 outburst of RS Ophiuchi. We detect extended emission in both \oiiil\ and \nevl\  lines. In both lines, the remnant has a double ring structure. The E-W orientation and total extent of these structures ($580 \pm 50$ AU at $d = 1.6$kpc) is consistent with that expected due to expansion of emitting regions imaged earlier in the outburst at radio wavelengths. Expansion at high velocity appears to have been roughly constant in the E-W direction ($v_{exp} = 3200 \pm 300$  km s$^{-1}$ in the plane of the sky), with tentative evidence of deceleration N-S. We present a bipolar model of the remnant whose inclination is consistent with that of the central binary. The true expansion velocities of the polar components are then $v = 5600 \pm 1100$  km s$^{-1}$. We suggest that the bipolar morphology of the remnant results from interaction of the outburst ejecta with a circumstellar medium that is significantly denser in the equatorial regions of the binary than at the poles. This is also consistent with observations of shock evolution in the X-ray and the possible presence of dust in the infrared. Furthermore, it is in line with models of the shaping of planetary nebulae with close binary central systems, and also with recent observations relating to the progenitors of Type Ia supernovae, for which recurrent novae are a proposed candidate. Our observations also reveal more extended structures to the S and E of the remnant whose possible origin is briefly discussed.

\end{abstract}





\section{Introduction}
\object{RS Ophiuchi} is a symbiotic recurrent nova (RN) which had previously undergone recorded outbursts in 1898, 1933, 1958, 1967 and 1985 \citep[see][]{ros87,rosi87}, with possible additional outbursts in 1907 \citep{sch04} and 1945 \citep{opp93}. On 2006 February 12.83UT it was observed to be undergoing a further eruption \citep{hir06}, reaching magnitude V=4.5 at this time. For the purposes of this paper, we define this as $t = 0$. The optical light curve then began a rapid decline, very similar to that seen in previous outbursts (\citealt{ros87}, AAVSO\footnote{http://www.aavso.org}). Several lines of evidence are consistent with a distance to RS Oph of $1.6\pm0.3$ kpc \citep{bod87}.

The RS Oph binary system comprises a red giant star in a $455.72\pm0.83$ day orbit with a white dwarf (WD) of mass near the Chandrasekhar limit \citep[see][]{dob94,sho96,fek00}. Accretion of hydrogen-rich material from the red giant onto the WD surface leads to the conditions for a thermonuclear runaway (TNR) in a similar fashion to that for classical novae (CNe). The much shorter inter-outburst period for this type of RN compared to CNe is thought to be due to a combination of the high WD mass and a supposed high accretion rate \citep[e.g.][]{sta85,yar05}. Such models lead to the ejection of somewhat lower masses at higher velocities than those for models of CN (typically 5000 km s$^{-1}$ and $10^{-8} - 10^{-6}$M$_\odot$ respectively for RNe). Spectroscopy of RS Oph has indeed shown H$\alpha$ line emission with FWHM $= 3930$ km s$^{-1}$ and FWZI $= 7540$ km s$^{-1}$ on 2006 February 14.2 ($t = 1.37$ days, \citealt{bui06}). 

Unlike CNe, where the mass donor is a low-mass main-sequence star, the presence of the red giant in the RS Oph system means that the high velocity ejecta run into a dense circumstellar medium in the form of the red giant wind, setting up a shock system with gas temperatures $\sim 2.2 \times 10^{8}$K for $v_{S} = 4000$ km s$^{-1}$, where $v_S$ is the velocity of the forward shock running into the pre-existing wind. X-ray observations of the 2006 outburst by \cite{bod06} and \cite{sok06} have confirmed the basic shock model of \cite{bod85} and \cite{obr92} in which RS Oph evolves like a supernova remnant (SNR), but on timescales around $10^5$ times faster.

VLBI imaging at 6cm and 18cm began 13.8 days after outburst \citep{obr06}. The initial image showed a partial ring of non-thermal (synchrotron) emission, of radius 13.8 AU at $d = 1.6$ kpc, consistent with emission from the forward shock. Subsequently, more extended lobes aligned E-W gradually emerged, with that in the East appearing first. This morphology was consistent with that derived from more rudimentary VLBI observations 77 days after the 1985 outburst by \cite{tay89}, apparently showing jets of emission. \cite{obr06} proposed a simple geometrical model comprising an expanding double-lobed structure, with the major axis perpendicular to the proposed plane of the central binary orbit.

Here we report Hubble Space Telescope ({\it HST}) observations of the expanding nebular remnant taken at $t = 155$d and compare them to the structures seen earlier in the radio. 
We then go on to explore remnant geometry by comparison to a bipolar model and thus deduce more about the circumstances of the outburst and the wider implications of these results.

\section{{\em HST} Observations and data reduction}


RS Oph was observed (Prop. ID 11004) in Director's Discretionary Time 155 days after outburst on 2006 July 17 using the Advanced Camera for Surveys (ACS) on the {\it HST}.
The 0.025\arcsec pixel$^{-1}$ High-Resolution Channel (HRC) was used for the observations combined with three narrow-band filters to isolate the \ha+\nii\ (filter F658N), \oiiil\ (F502N) and \nevl\ (F343N) nebular emission lines. In order to reduce readout overheads and maximise on-object time, the 512x512 square pixel subarray was used throughout. 

One orbit was spent imaging RS Oph, while a second was used to observe HD 166215, a bright nearby standard star of comparable spectral type (V = 9.32, K0). With limited time for the RS Oph observations, a balance was struck between deep imaging and dithering to increase resolution. Care was taken when planning the exposures not to saturate either the primary target or standard star images in any filter. The resulting images were checked and verified as unsaturated in every case. Observations of RS Oph were repeated with a 3\arcsec\ offset in both RA and declination in order to identify and eliminate ghosts if present. A two point dither pattern was used for each observation. Table 1 summarises the resulting observations of each target.

All data were reprocessed using standard procedures outlined in the ACS Data Handbook\footnote{http://www.stsci.edu/hst/acs/documents/handbooks/DataHandbookv4/ACS\_longdhbcover.html} and the Pydrizzle\footnote{http://stsdas.stsci.edu/pydrizzle} and Multidrizzle\footnote{http://stsdas.stsci.edu/pydrizzle/multidrizzle} Handbooks, using optimal input parameters to maximise the signal to noise ratio of each image. Although extended emission was clearly visible in the unprocessed \oiii\ images, a  number of data analysis techniques were used in order to explore in detail the presence of extended emission in the images for each filter.

PSF profiles generated by {\it TinyTim} \citep{kri95} were then subtracted from the RS Oph data in each of the emission lines. The PSF was scaled to an extent such that when subtracted, it brought the formerly peak pixel value to approximately the same level as those surrounding it. After subtraction, the total counts over an area enclosing the central star (as a percentage of the pre-subtracted values) are 92\% (\oiii\ and \nev) and 65\% (\ha). It is likely therefore that there is some residual stellar flux in the central regions of the PSF-subtracted images.

Using the profile of HD 166215 as the PSF, and also that generated by {\it TinyTim}, deconvolution using the Lucy-Richardson method was performed on the RS Oph data for each emission line. Tests using both the CLEAN and  Maximum Entropy techniques produced similar results. However, for \nev\, the lower signal-to-noise of the stellar PSF meant that only the {\it TinyTim} PSF could be used effectively in the deconvolution. 


As part of this study, a re-analysis was carried out of pre-outburst WFPC2 observations of the \oiiil\ line through the 502N filter on 2000 June 12 (Prop. ID 8332). No extended emission, even at very faint levels, was detectable. This confirms the results of the analysis of these data undertaken by \cite{bro03}.


\section{Results and Discussion}

\begin{figure}
\epsfclipon
\centering
\mbox{\epsfxsize=6.4in\epsfbox[0 0 414 414]{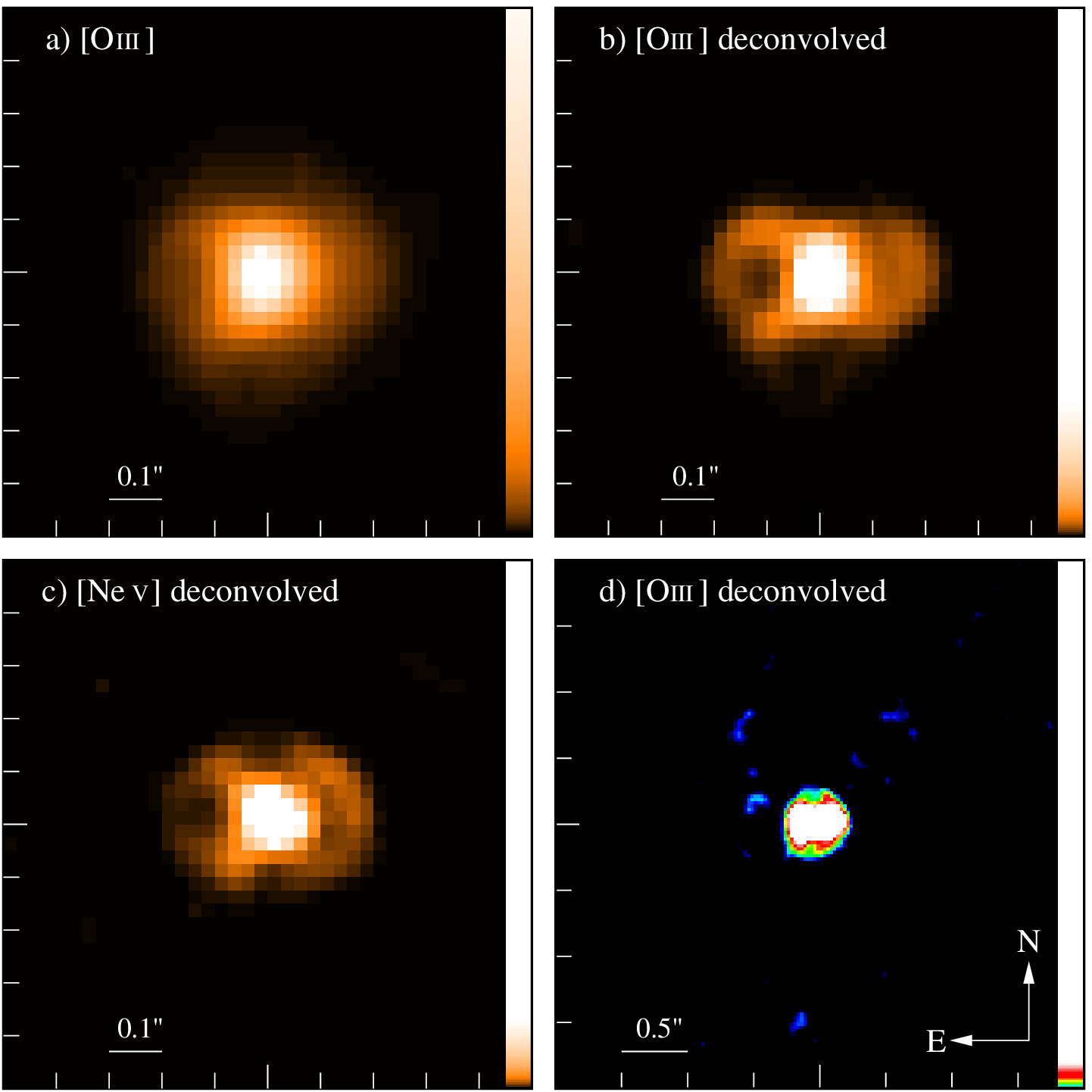}}
\caption{
HST ACS/HRC narrow band images of RS Oph (a) PSF-subtracted \oiiil\ image clearly showing extended emission at sub-arcsecond size scales, particularly in the E-W direction (note that the PSF-subtracted \nevl\ image also showed clear evidence of extended emission); (b) results of deconvolution of the RS Oph PSF-subtracted image with the PSF star, again in \oiiil, showing a double ring structure; (c) as for (b), but for \nevl\ and using a PSF generated by {\it Tiny Tim}; (d) deeper, larger area view of (b), showing an arc-like feature to the E and a southern ``blob'' of emission (see text for further details). Surface brightness is shown on a logarithmic scale on all images, with the vertical bar showing linear representation of the color table from 0 to 50\% of the peak emission.\label{images}
}
\end{figure}

As can be seen in Figure~\ref{images}, extended structure was clearly visible in the \oiiil\ line in the PSF-subtracted image (and was indeed visible in the raw image). Deconvolution revealed more detailed structure in both \oiii\ and \nevl. There was also a hint of possible extended emission close to the central source in the H$\alpha$ line, but this was not present at a significant level.

In the deconvolved \oiii\ and \nev\ images, the most striking feature is a double ring structure with major axis lying E-W and total (peak-to-peak) extent $360 \pm 30$ mas ($580 \pm 50$ AU at $d = 1.6$ kpc). The most extended structures seen in the radio \citep[the outer lobes;][]{obr06} lie along this axis. Assuming ejection at the time of the outburst, the expansion rate (from the center) of the optical emission along this axis is $1.2 \pm 0.1$ mas d$^{-1}$ (equivalent to $v_{exp} = 3200 \pm 300$  km s$^{-1}$ in the plane of the sky). The optical emission is also detectable above background in the deconvolved images to a total extent of $520 \pm 50$ mas, corresponding to an expansion rate of $1.7 \pm 0.2$ mas d$^{-1}$. We may compare this to 1.4 $\pm0.3$ mas d$^{-1}$ for the E-W lobes seen in the radio, taken over 4 epochs from $t = 21.5$ to 62.7 days in the 2006 outburst (O'Brien et al., in preparation) and 1.3 mas d$^{-1}$ for the equivalent features derived from VLBI observations in the 1985 outburst \citep{tay89}. Thus there is evidence that the bipolar emission seen here and in the radio arises from the same regions of the remnant, if the expansion velocities in the E-W direction are roughly constant over the first 155 days.

Tracing the ring-like structure of the optical remnant suggests a N-S (peak-to-peak) extent of $150 \pm 25$ mas. This corresponds to an expansion rate of $0.48 \pm 0.08$ mas d$^{-1}$ compared with $0.62 \pm 0.01$ mas d$^{-1}$ derived in the radio for the central ring on day 13.8 \citep{obr06}. As noted in \cite{obr06}, the formal error on this radio expansion rate is likely to be an underestimate; \cite{rup07} also derived radio expansion rates for the ring $\sim20$\% higher. Thus comparison of the HST and radio data tentatively suggests deceleration in the N-S direction, but verification awaits further optical observations.


As noted above, both \oiii\ and \nev\ show similar structure. In the case of both lines, emission might arise either from regions downstream of the forward shock in the red giant wind or from a ``precursor'' ionised region ahead of the shock, or both \citep[e.g.][]{dop96}. The more marginal presence of \ha+\nii\ may be a direct consequence of particular shock velocities \citep[again see e.g.][]{dop96} coupled with decreasing contrast of extended emission with the central star at longer wavelengths. It is also of interest to note that the assumption from $Spitzer$ observations of infrared fine structure and coronal lines \citep{eva07} of an origin of these lines in the same remnant regions is consistent with our observations.

\begin{figure}
\epsfclipon
\centering
\mbox{\epsfxsize=6.4in\epsfbox[0 0 414 414]{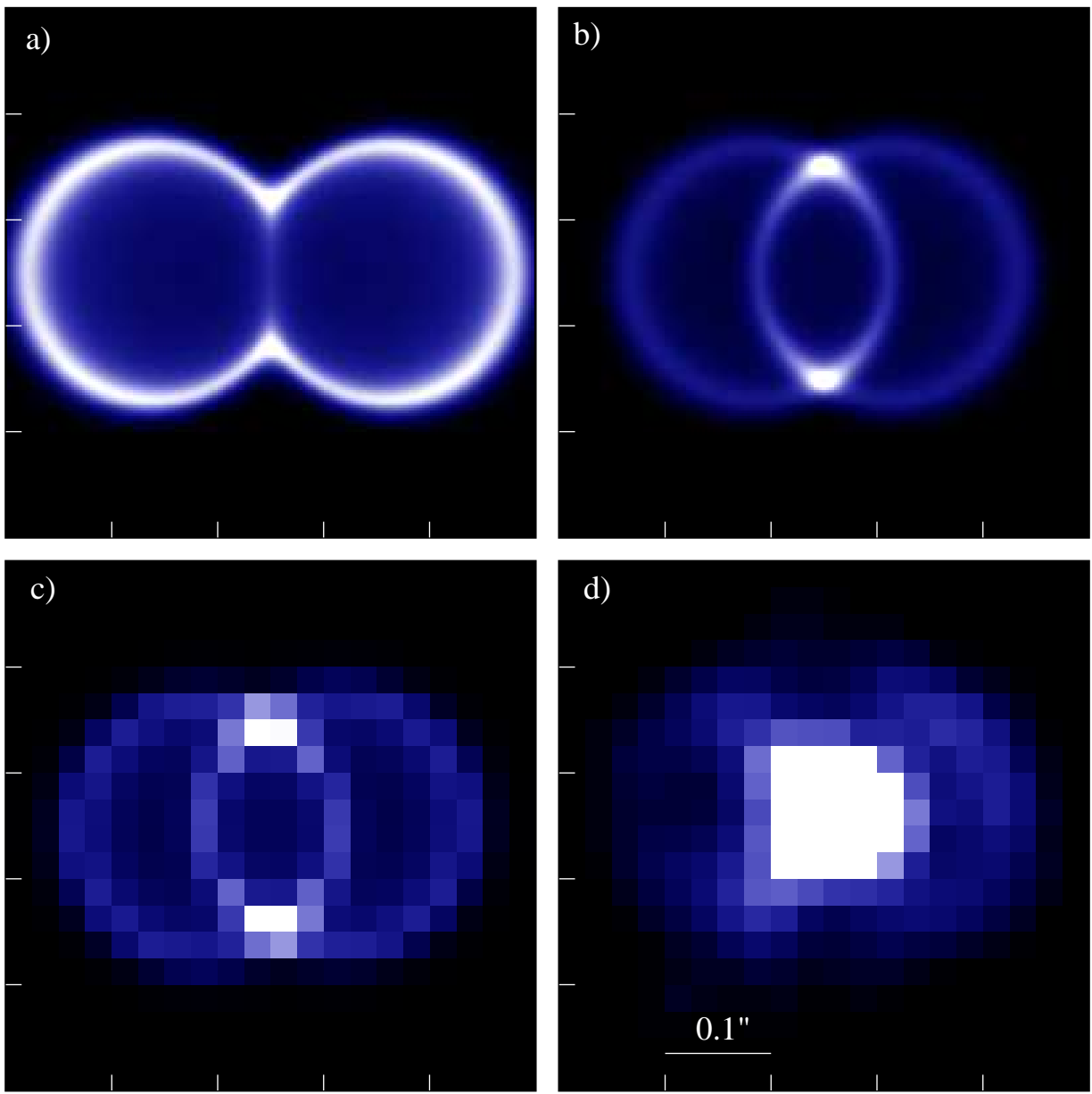}}
\caption{
Comparison of the \oiiil\ image with a simple emission model. (a) Side-on view of the bipolar nebula, with finite shell thickness. Here, the plane of the central binary system lies along the ``waist'' of the remnant. Surface brightness is simply proportional to line of sight path length through the shell; (b) Model inclined by $35\degree$ to the line of sight, consistent with the inclination of the orbit from orbital solutions \citep{dob94}; (c) Resulting model with resolution degraded and image rotated $3\degree$ anticlockwise from an E-W axis to match that of the ACS/HRC \nevl\ image shown in (d) for comparison. \label{model}
}
\end{figure}

\cite{obr06} proposed a simple model for the radio emission comprising a bipolar structure where the evolution of the optical depth due to free-free absorption led to the gradual uncovering of various features, most notably this explained the emergence of the outermost radio lobes. Here, we have modeled the optical emission seen in our HST images with a ``peanut-shaped'' bipolar structure using the modeling code described in \cite{har03} (see Figure~\ref{model}). The shell has finite thickness ($0.05$ arcsec) and the surface brightness distribution is just determined by the line-of-sight path length through the nebula. The axial ratio is set at 3:1, consistent with a constant velocity of expansion along the major axis and deceleration along the ``waist'', as implied by the earlier radio observations.


As can be seen from Figure~\ref{model}, the model reproduces the morphology of the optical emission extremely well for an inclination $i = 35\degree$, consistent with the major axis lying normal to the plane of the binary orbit, where $i = 30\degree - 40\degree$ was given by \cite{dob94} from radial velocity studies. Following \cite{obr06}, we also note that the 2006 outburst occurred at approximately the same binary phase as in 1985, and hence the similarity between the rate of expansion of the radio lobes on the plane of the sky in the two outbursts would naturally follow.



We now turn our attention briefly to the features to the E and S of the main remnant (Figure~\ref{images}). The eastern ``arc'', which is apparent only in the \oiii\ image, has no counterpart detectable to the W and its closest approach to the centre of the bipolar emission is $0.44 \pm 0.01$ arcsec. Similarly, the southern ``blob'' (also detected in \ha) lies at $1.52 \pm 0.02$ arcsec from the centre. If these features arose in the 2006 outburst, for $d = 1.6$ kpc, they would therefore imply $v_{exp}\sim8000$ and $\sim27000$ km s$^{-1}$ respectively, values far higher than those otherwise observed or expected in such an outburst. On the other hand, at this time, one might expect light echoes in the surrounding medium to have radial extent far greater than this ($\sim16.3$ arcsec, for a scattering medium in the plane of the sky at the distance of the nova). Another possibility is that these features are in fact associated with material piling up from previous outbursts. Future HST observations will help to clarify their origin, and also explore the late-time evolution of the remnant.



\section{Conclusions and future work}

We have detected, in at least two emission lines, the resolved optical counterparts of the expanding emission first seen much earlier in the outburst at radio wavelengths. The central remnant is then modeled in terms of a bipolar structure whose inclination is consistent with that of the binary orbit. For $d = 1.6 \pm 0.3$ kpc \citep{bod87}, $i = 35\degree$, an expansion velocity, $v = 5600 \pm 1100$ km s$^{-1}$ and $v_{rad} = 4600 \pm 900$ km s$^{-1}$ are implied for the material at the poles (the formal errors being dominated by the uncertainty in distance here). Unless the ejection of material at the explosion site is highly anisotropic (which appears unlikely) this reinforces the notion that remnant shaping occurs due to the interaction of the ejecta with the pre-existing circumstellar environment. The implication would then be that the densest parts of the red giant wind lie in the equatorial regions along the plane of the binary orbit. This in turn leads to more marked deceleration of material in this direction compared to the poles. The existence of a dense region of the wind, which has not been fully traversed by a forward shock (even at these relatively late times) may well explain the apparent presence of dust in the circumstellar environment throughout the outburst (Evans et al., in preparation). It is from this dense equatorial region that we now suspect most of the early X-ray emission also arises as the forward shock driven into this region by the outburst ejecta decelerates. The rapid evolution of the remnant derived by \cite{bod06} is consistent with this proposal.

These observations have wider relevance to our understanding of the shaping of nova shells and planetary nebulae 
For example, \cite{mit07} have shown that the PN Abell 63 comprises narrow lobes which form a ``tube'' at the ends of which are ``caps'' of emission. At the centre of Abell 63 is the eclipsing close binary UU Sge \citep[$p = 11.2$ hrs][]{bon78}, and \cite{mit07} have demonstrated that the inclination of the lobes is that of the plane of the central binary. They conclude therefore that the PN was shaped by interaction during its early evolution with a circumbinary environment with enhanced density in the equatorial plane. This is consistent with the results of hydrodynamic modelling of the evolution of the environment of such systems performed by \cite{mas99}. Although they did not extend their study to semi-detached (rather than detached) systems, these authors found that equatorial enhancements in the density of the winds from red giants and AGB stars in binary systems were most pronounced for those with the smallest binary separations (i.e. comparable to that in RS Oph).

Recent observations by \cite{pat07} of the Type Ia SN 2006X show spectral features that may be associated with the interaction of the SN ejecta with multiple, slow moving shells of pre-existing material, which have been suggested to arise from a recurrent nova progenitor system. Furthermore, this interpretation is aided by an anisotropic density distribution in the pre-supernova circumstellar environment, compatible with enhanced density in the equatorial plane of the progenitor binary system.

Finally, it is clear that our current spherically-symmetric hydrodynamic models of the evolution of shock structures in RS Ophiuchi will need revision to take account of the geometry revealed in these observations. Such modeling is now underway (Vaytet et al., in preparation). Future work also includes the analysis of a further planned epoch of HST observations to explore more details of remant evolution and which may in addition elucidate the nature of the more extended features seen in our day 155 observations. In a future paper, we hope to combine all our HST observations with contemporaneous high resolution optical spectroscopy to enable us to produce the most realistic models to-date of the true geometry of the remnant.


\acknowledgments

The authors are very grateful to the {\it HST} Director for provision of Discretionary Time and the {\it HST} support staff, in particular Patricia Royle and Kailash Sahu, for their assistance with planning and analysing the observations reported here. We thank Philipp Podsiadlowski and Stephen Justham for very informative discussions on the circumbinary environment and the link to Type Ia supernovae, and Shazrene Mohamed who directed us to the Mastrodemos and Morris paper. We are also grateful to Michael Shara for helpful discussions regarding our HST program and an anonymous referee for helping to improve the original manuscript. SS gratefully acknowledges partial support from NSF and NASA grants to ASU. SPSE acknowledges the support of the UK's Nuffield Foundation. MFB was supported by a PPARC Senior Fellowship. 



{\it Facilities:} \facility{HST}




\pagebreak

\begin{table}
\centering
\begin{tabular}{lll}\hline
Object         & Filter & Exp Time (s) \\\hline
RS Oph         & F343N  & 240          \\
RS Oph         & F502N  & 180          \\
RS Oph         & F658N  & 40           \\
HD166215       & F343N  & 498          \\
HD166215       & F502N  & 40          \\
HD166215       & F658N  & 16          \\
\hline
\hline
\end{tabular}
\caption{Summary of total exposure times per filter on both target objects observed with the HST ACS/HRC in Program 11004 on 2006 July 17.}
\end{table}

\clearpage

\end{document}